\documentclass[twocolumn,showpacs,superscriptaddress,amsmath,amssymb]{revtex4-1}
\usepackage{graphicx}
\usepackage{dcolumn}
\usepackage{bm}
\usepackage{color}
\usepackage{epsfig}
\usepackage{epstopdf}
\usepackage{mathrsfs}
\usepackage{hyperref}
\usepackage{amsmath}
\usepackage{amsxtra}
\usepackage{amsfonts}

\usepackage{multirow}
\usepackage{makecell}
\usepackage{longtable}
\usepackage{diagbox}
\usepackage{slashbox}

\def\bd{
\begin{document}} \def\ed{\end{document}}
\def\bmp{\begin{minipage}} \def\emp{\end{minipage}}
\def\bcc{\begin{center}} \def\ecc{\end{center}}     \def\npg{\newpage}
\def\beq{\begin{equation}} \def\eeq{\end{equation}} \def\hph{\hphantom}
\def\be{\begin{equation}} \def\ee{\end{equation}} \def\r#1{$^{[#1]}$}
\def\n{\noindent} \def\ni{\noindent} \def\pa{\parindent}
\def\hs{\hskip} \def\vs{\vskip} \def\hf{\hfill} \def\ej{\vfill\eject}
\def\cl{\centerline} \def\ob{\obeylines}  \def\ls{\leftskip}
\def\underbar#1{$\setbox0=\hbox{#1} \dp0=1.5pt \mathsurround=0pt
   \underline{\box0}$}   \def\ub{\underbar}    \def\ul{\underline}
\def\f{\left} \def\g{\right} \def\e{{\rm e}} \def\o{\over} \def\d{{\rm d}}
\def\vf{\varphi} \def\pl{\partial} \def\cov{{\rm cov}} \def\ch{{\rm ch}}
\def\la{\langle} \def\ra{\rangle} \def\EE{e$^+$e$^-$} \def\pt{p_{\rm t}}
\def\pti{p_{{\rm t},i}} \def\vti{v_{{\rm t},i}}
\def\ptj{p_{{\rm t},j}}\def\Pt{P_{\rm t}} \def\vt{v_{\rm t}}

\def\bitz{\begin{itemize}} \def\eitz{\end{itemize}}
\def\btbl{\begin{tabular}} \def\etbl{\end{tabular}}
\def\btbb{\begin{tabbing}} \def\etbb{\end{tabbing}}
\def\beqar{\begin{eqnarray}} \def\eeqar{\end{eqnarray}}
\def\\{\hfill\break} \def\dit{\item{-}} \def\i{\item}
\def\bbb{\begin{thebibliography}{9} \itemsep=-1mm}
\def\ebb{\end{thebibliography}} \def\bb{\bibitem}
\def\bpic{\begin{picture}(260,240)} \def\epic{\end{picture}}
\def\akgt{\cl{\bf ACKNOWLEDGMENTS}}
\def\fgn{\noindent{\bf\large\bf figure captions}}
\def\m1{\langle N_p\rangle} \def\u2{\langle N_{\bar p}\rangle} \def\Nap{N_{\bar
p}}
\def\lan{\langle}
\def\ran{\rangle}
\def\p{\pi}
\def\ifmath#1{\relax\ifmmode #1\else $#1$\fi}%
\def\rc{\ifmath{{\mathrm{c}}}}
\def\cut{\ifmath{{\mathrm{cut}}}}
\def\rF{\ifmath{{\mathrm{F}}}}
\def\rK{\ifmath{{\mathrm{K}}}}
\def\rp{\ifmath{{\mathrm{p}}}}
\def\rt{\ifmath{{\mathrm{t}}}}
\def\LAB{\ifmath{{\mathrm{LAB}}}}
\def\cut{\ifmath{{\mathrm{cut}}}}
\def\beq{\begin{equation}}
\def\eeq{\end{equation}}

\newcommand{\cinst}[2]{$^{\mathrm{#1}}$~#2\par}
\newcommand{\crefi}[1]{$^{\mathrm{#1}}$}
\newcommand{\crefii}[2]{$^{\mathrm{#1,#2}}$}
\newcommand{\crefiii}[3]{$^{\mathrm{#1,#2,#3}}$}
\newcommand{\HRule}{\rule{0.5\linewidth}{0.5mm}}
\newcommand{\minitab}[2][l]{\begin{tabular}{#1}#2\end{tabular}}

\bd

\title{Subtracting non-critical fluctuations from higher cumulants of conserved charges by the sample of mixed events}

\author{Fan Zhang}
\affiliation{Key Laboratory of Quark and Lepton Physics (MOE) and
Institute of Particle Physics, Central China Normal University, Wuhan 430079, China}
\author{Zhiming Li}
\affiliation{Key Laboratory of Quark and Lepton Physics (MOE) and
Institute of Particle Physics, Central China Normal University, Wuhan 430079, China}
\author{Lizhu Chen}
\affiliation{School of Physics and Optoelectronic Engineering, Nanjing University of Information Science and Technology, Nanjing 210044, China}
\author{Xue Pan}
\affiliation{School of Electronic Engineering, Chengdu Technological University, Chengdu 611730, China}
\author{Mingmei Xu}
\affiliation{Key Laboratory of Quark and Lepton Physics (MOE) and
Institute of Particle Physics, Central China Normal University, Wuhan 430079, China}
\author{Yeyin Zhao}
\affiliation{Key Laboratory of Quark and Lepton Physics (MOE) and
Institute of Particle Physics, Central China Normal University, Wuhan 430079, China}
\author{Yu Zhou}
\affiliation{Department of Physics and Astronomy University of California Los Angeles, CA 90095, USA}
\author{Yuanfang Wu}\email{wuyf@mail.ccnu.edu.cn}
\affiliation{Key Laboratory of Quark and Lepton Physics (MOE) and
Institute of Particle Physics, Central China Normal University, Wuhan 430079, China}

\begin{abstract}
It is generally known that the sample of mixed events keeps global and systematic characters of the original sample, and omits concerning correlations. To subtract non-critical fluctuations in the sample of mixed events, we define dynamical cumulants of conserved charges. Using the sample produced by the AMPT default model, we construct its corresponding sample of mixed events, and demonstrate that dynamical cumulants reduce influences of statistical fluctuations, and non-critical effects in the sample of mixed events. They are also centrality bin width and detection efficiency independent, in consistent with those of formulae corrected cumulants. Therefore, dynamical cumulants present more critical related fluctuations. 
\end{abstract}


\maketitle
\section{Introduction}
To map QCD phase diagram from the experimental side, programs of beam energy scan (BES) I and II at the relativistic heavy-ion collider (RHIC) are proposed and in progress~\cite{scan-a,scan-b,scan-c,scan-d,scan-e,BES2}. Higher cumulants of conserved charges are suggested observables, which are sensitive to critical fluctuations~\cite{m1,m2,m3,m4,m5,m6,m7,m8,m9,m10,m11,m12,m13}. At the RHIC BES I, non-monotonic event-by-event fluctuations of net-proton cumulants have been found~\cite{star-prl1,star-prl2}. Whether they are critically related fluctuations are highly interested, and crucial for locating the critical point (CP) of QCD phase transition at the current RHIC BES II~\cite{BES2}. 

As we have known, correlation length in vicinity of the critical point is divergent for infinite system size, and large for finite size. In heavy-ion collisions, critically related fluctuations should also be large, and come from the inner correlations of an event. Higher cumulants measured event-by-event fluctuations of conserved charges, and are closely related to critical fluctuations. 

However, non-critical fluctuations also contribute to cumulants, and change with incident energy and impact parameter. Non-critical fluctuations are mainly two sources. One is conventional mechanisms, such as the resonance decay, and global conservation of energy, momentum, and various charges. The length of these conventional correlations is finite and fixed. Their fluctuations are usually small in comparison to critical one~\cite{Songhc-PRC}.
 
Another source is global and systematic effects, such as the statistical fluctuations due to insufficient number of particles~\cite{Panx-PRC, Zhoudm-PRC}, initial size fluctuations for different impact parameters~\cite{Songhc-PRC,Songhc-PRC2,Songhc-Npa}, centrality bin width~\cite{Luoxf-CBWC}, detection efficiency and acceptance cuts~\cite{Koch-PRC91-027901}, and so on. 

Up to now, numerous efforts have been made to eliminate non-critical effects~\cite{global,Luoxf-CBWC, Panx-PRC, Zhoudm-PRC, Koch-PRC91-027901, Songhc-PRC}. For example, statistical fluctuations are usually estimated by corresponding Poisson distribution~\cite{Lizhu-JPG1,Lizhu-JPG2,Panx-PRC}. To reduce the influence of centrality bin width, a well known scheme, centrality bin width correction (CBWC), is proposed~\cite{Luoxf-PRC}. To eliminate the influence of detection efficiency, a complex formula which connects the true cumulant to that of actually measured is introduced~\cite{Koch-PRC-efficiency,Koch-PRC91-027901}.

For a real world sample, all kinds of non-critical effects are involved. Correlations caused by conventional machanisms are difficult to separate from the critical one. While, those global and systematic effects can be well taken together into account by so called {\it the sample of mixed events}~\cite{mix,Zhoudm-PRC}, where all concerning correlations are removed. So cumulants of mixed sample provide an estimation for most background effects.

To subtract the contribution of the mixed sample, we define dynamical cumulant as the difference between cumulants of original and mixed samples~\cite{dynamical-1,dynamical-2,dynamical-3,Lizhu-JPG1,Lizhu-JPG2,Panx-PRC}. They should be more critically related fluctuations. 
 
In this paper, we first introduce the method of mixed events in Section II. Then in section III, using the sample of Au + Au collisions at 19.6 GeV produced by a multiphase transport (AMPT) default model, we construct its corresponding sample of mixed events. Higher cumulants of net-proton for the original and mixed samples are calculated and compared. Influences of statistical fluctuations, centrality bin width, and efficiency are studied respectively. Finally, a brief summary and conclusions are given in section IV. 

\section{Construction of mixed sample}

Usually, the method of mixed events changes with observable. For higher cumulants of conserved charges, they are defined as variance ($\sigma^2$), skewness ($S$), kurtosis ($\kappa$), and their products, $S\sigma$ and $\kappa\sigma^2$, i.e.,
\begin{equation}\label{cumulants}
\begin{split}
\sigma^2 &=\langle(\triangle N_{\rm c})^2\rangle, \\
S &=\langle(\triangle N_{\rm c})^3\rangle/\sigma^3,\\
\kappa &=\langle(\triangle N_{\rm c})^4\rangle/\sigma^4-3,\\
S\sigma &=\langle(\triangle N_{\rm c})^3\rangle/\sigma^2,\\
\kappa\sigma^2 &=\langle(\triangle N_{\rm c})^4\rangle/\sigma^2-3\sigma^2.
\end{split}
\end{equation} 
Where the average $\langle\rangle$ is over the whole event sample. $\triangle N_{\rm c}=N_{\rm c}-\langle N_{\rm c}\rangle$. $N_{\rm c}$ is the number of particles with conserved charges, which usually refer to baryon, strangeness, or electric charge. In this paper, we restrict the conserved charge to baryon, or strangeness only. The total number of particles in an event is electric charged, i.e, $N_{\rm ch}$,  multiplicity. 

By definition, Eq.~(\eqref{cumulants}), $N_{\rm c}$ is correlated with its associated event. Charged particles in a given event correlate with each other and with $N_{\rm ch}-N_{\rm c}$ particles as well. For a mixed event, all such correlations have to be removed. Meanwhile, the global characters, such as, the multiplicity $N_{\rm ch}$ distribution, and the mean number of charged particles $\langle N_{\rm c} \rangle$, which directly relate to initial size and statistical fluctuations, detection efficiency and acceptance cuts, should retain. 

We have demonstrated that for cumulates of conserved charges, the pool method~\cite{Zhangf-mixed} is effective in constructing mixed events. Where multiplicity $N_{\rm ch}$ is simply taken from the original sample. All particles of events is firstly put into a pool, and then randomly take $N_{\rm ch}$ particles from the pool to form a mixed event. If the sample is large enough, i.e., $N_{\rm event} \gg N_{\rm ch}$, $N_{\rm ch}$ particles should come approximately from different events and the correlations among particles are negligible. Meanwhile, it has shown that the mean number of charged particles $\langle N_{\rm c} \rangle$ is consistent with that of the original sample.

For a given sample, its corresponding mixed events can be constructed by the pool method. For convenience in following statements, we label cumulants of the original and mixed samples by superscripts $o$ and $m$, respectively. We define dynamical cumulant as the difference between cumulants of the original and mixed samples, i.e., 
\begin{equation}\label{dyn cumulants}
\begin{split}
\ \sigma^{2}_{\rm dyn} &=\sigma^{2,\rm o}-\sigma^{2,\rm m}, \\
 S_{\rm dyn} &=S^{\rm o}-S^{\rm m},\\
 \kappa_{\rm dyn}&=\kappa^{\rm o}-\kappa^{\rm m},\\
 S\sigma_{\rm dyn} &=(S\sigma)^{\rm o}-(S\sigma)^{\rm m},\\
\kappa\sigma^2_{\rm dyn}&=(\kappa\sigma^2)^{\rm o}-(\kappa\sigma^2)^{\rm m}.
\end{split}
\end{equation} 

Obviously, non-critical effects contained in the mixed sample are subtracted by defined dynamical cumulants~\cite{Lizhu-JPG1,dynamical-1,dynamical-2,dynamical-3}. In the following, we will demonstrate the effectiveness of dynamical cumulants in subtracting statistical fluctuations, and compare dynamical cumulants with those of formulae corrected, $\sigma^2_{\rm fc}$, $S_{\rm fc}$, $ \kappa_{\rm fc}$, $ S\sigma_{\rm fc}$, and $\kappa\sigma^2_{\rm fc}$. 

\section{Application and comparison}

We generate a sample of Au + Au collisions at 19.6 GeV by the AMPT default model. 
Where multiplicity is total electric charged particles, i.e., $N_{\rm ch}$, and the conserved charge is considered approximately by proton~\cite{net-proton1,net-proton2,net-proton3}. Cuts in transverse momentum and rapidity are $ 0.4 \le p_t \le 0.8$ (GeV) and $ -0.5\le y \le 0.5$, the same as those used at the RHIC/STAR experiments~\cite{Star-centrality}. At each of centrality bins, corresponding sample of mixed events is constructed by the pool method.

As described in the Introduction Section, up to now, there are only three kinds of formula corrected cumulants, which correctted influences of statistical fluctuations~\cite{Lizhu-JPG1, Panx-PRC, Zhoudm-PRC}, centrality bin width and detection efficiency, respectively~\cite{Luoxf-PRC91-034907, Koch-PRC91-027901}. So in the following, we will study them respectively, and compare dynamical cumulants with those of formula corrected. 

\subsection{Statistical fluctuations}

Usually, statistical fluctuations are described by a Poisson-like distribution with $\langle N_{\rm c} \rangle$ of the original sample~\cite{Lizhu-JPG1,Panx-PRC,Zhoudm-PRC}. The formula corrected cumulant can be considered as that of subtracting Poisson distribution. 

The centrality dependences of $S\sigma$ and $\kappa\sigma^2$ of original (black circles) and mixed (red open squares) samples are presented in Fig.~1(a) and (c). Where 9 centrality bins are defined by multiplicity distribution, same as those given at the RHIC/STAR~\cite{Star-centrality}. The centrality dependence of dynamical (violet open crosses) and formula corrected (blue open diamonds) $S\sigma$ and $\kappa\sigma^2$ are presented in Fig.~1(b) and (d), respectively.
\begin{figure}
\includegraphics[width=.8\linewidth,angle=-90]{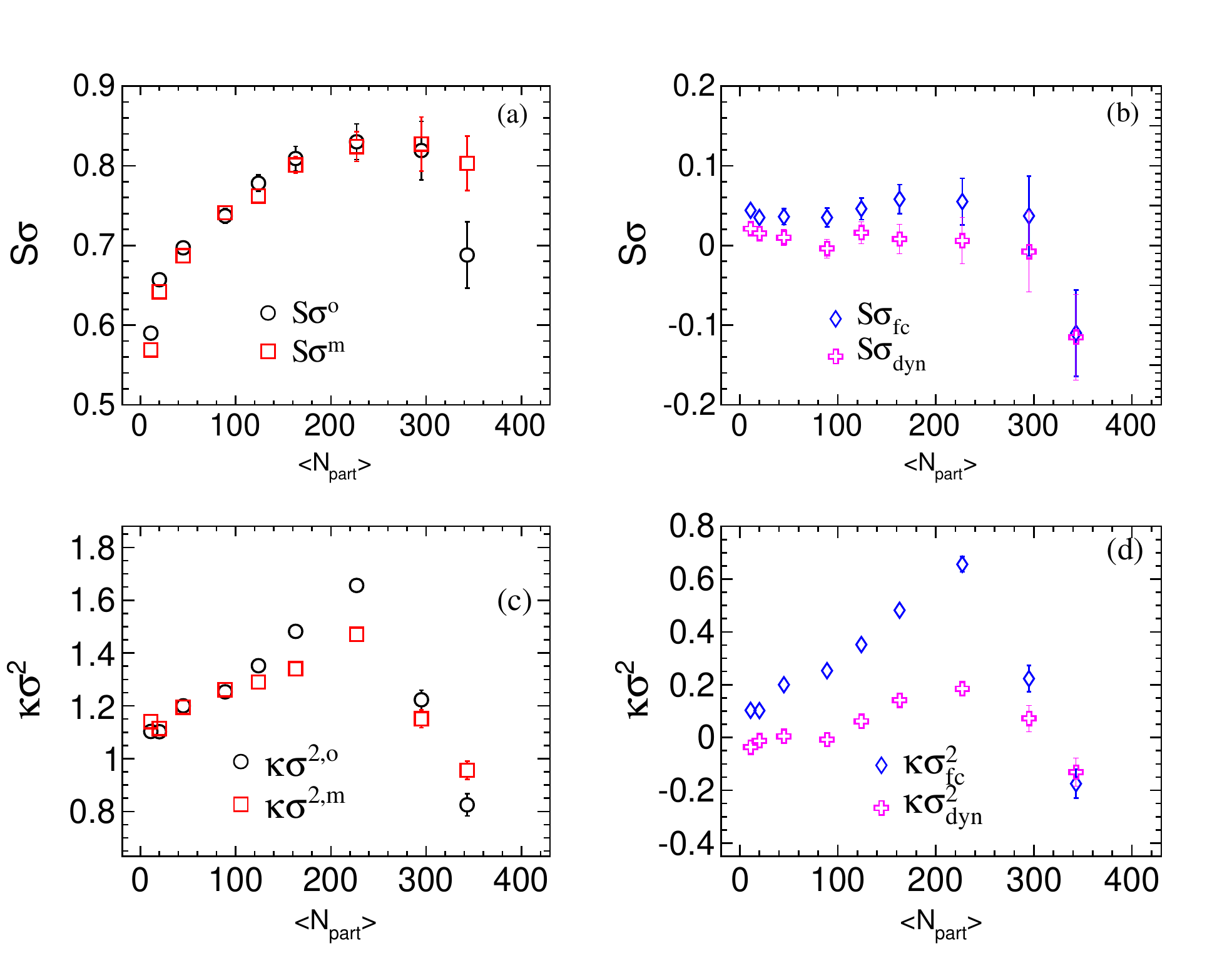}
\caption{\label{Fig. 1}Left:  the centrality dependence of $S\sigma$ (a) and $\kappa\sigma^2$ (c) for the original (black circles) and mixed (red open squares) samples. Right: the centrality dependence of dynamical (violet open crosses) and formula corrected (blue open diamonds) $S\sigma$ (b) and $\kappa\sigma^2$ (d).}
\end{figure}

Fig.~1(a) shows that $S\sigma$ of mixed sample, i.e., red open squares,  are close to those of the original sample, i.e.,  $S\sigma^{\rm o}$, black circles, except one for the most central collision. In Fig.~1(b),  most of $S\sigma_{\rm dyn}$, violet open crosses, are flat and show a low shift to those of $S\sigma_{\rm fc}$, blue open diamonds. Only for the most central collision, the violet open crosses almost overlaps with the blue open diamonds.

Differences between $S\sigma_{\rm dyn}$ and $S\sigma_{\rm fc}$ are understandable. As we have discussed in the Introduction Section, dynamical cumulants deduct most of global and systematic effects remaining in the mixed sample. Here, formula corrected cumulant only deducted Poisson-like statistical fluctuations. Other global effects, in particular, the initial size fluctuations remain. So formular corrected cumulant is larger than dynamical one. 

Fig.~1(c) also shows that the general trend of the centrality dependence of $\kappa\sigma^2$ of the mixed sample is analogous with that of the original sample. Differences between them are visible, or larger, in comparison to the case of $S\sigma$ in Fig.~1(a). In Fig.~1(d), the centrality dependence of $\kappa\sigma^2_{\rm dyn}$ is qualitatively similar to that of $\kappa\sigma^2_{\rm fc}$, but much smooth and lower than that of $\kappa\sigma^2_{\rm fc}$. 


So in general, dynamical cumulant is smaller than that of formula corrected, where only Poisson-like statistical fluctuations are subtracted. This implies that mixed sample contains other non-critical effects in addition to Poisson-like statistical fluctuations. We will show in the following that dynamical cumulants also eliminate influences of centrality bin width and detection efficiency. 

Meanwhile, it should be noted although dynamical cumulants showed in Fig.~1(b) and (d) are very small, but not zero. They present conventional correlations implemented in the AMPT model. 

\subsection{Centrality bin width}

It is known that a given centrality bin width composites a superposition of various impact parameters. 
To study the influence of centrality bin width, we firstly present the centrality dependence of $\kappa\sigma^2$ of the original sample for nine (black solid points) and sixteen (red circles) centrality bins in Fig.~2(a). It shows clearly that $\kappa\sigma^2$ decreases with the bin width, as expecting~\cite{Luoxf-CBWC}. 

\begin{figure}
\includegraphics[width=1.0\linewidth,angle=-90]{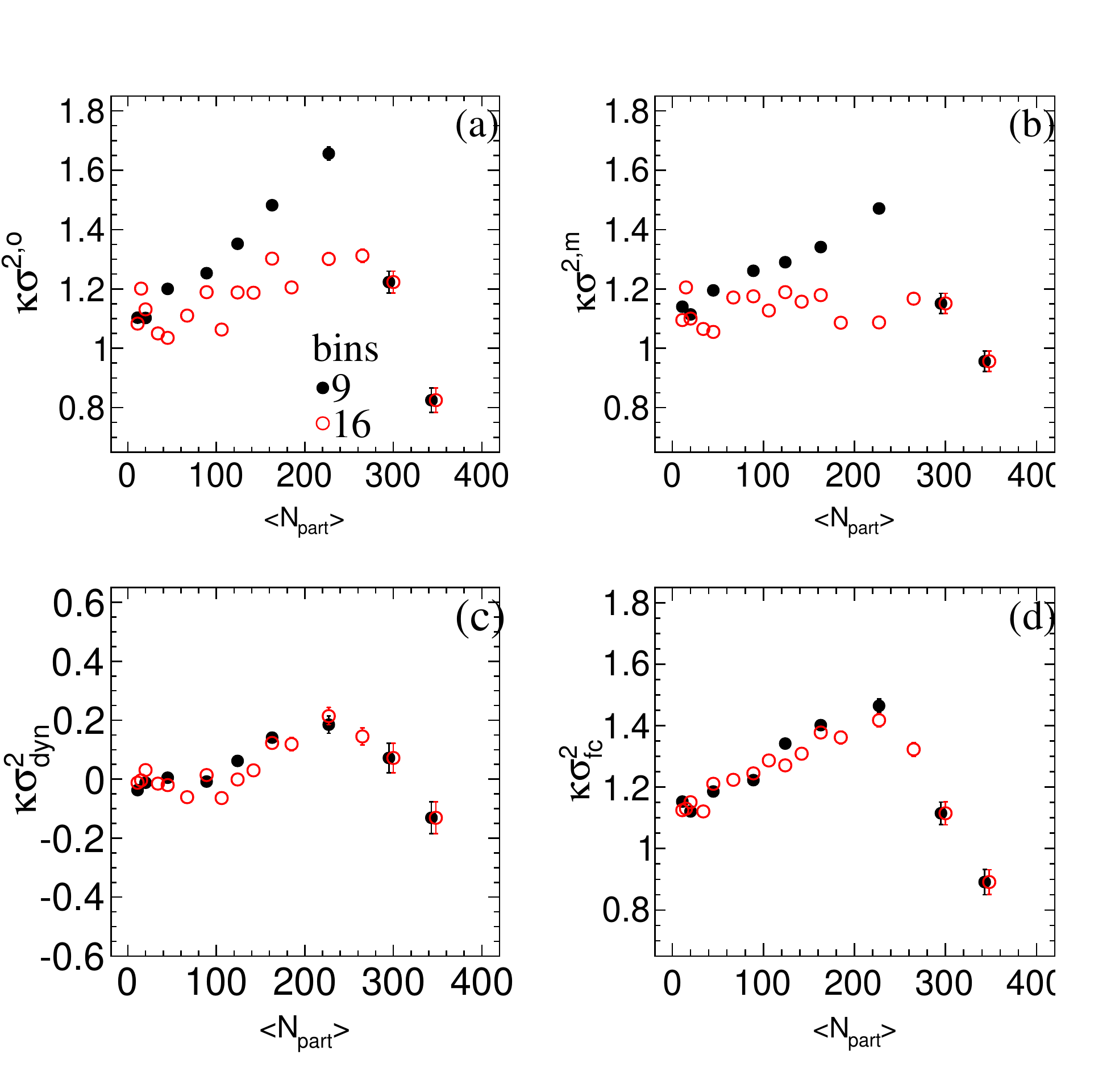}
\caption{\label{Fig. 2}The centrality dependence of $\kappa\sigma^2$ of original (a) and mixed (b) samples, and $\kappa\sigma^2_{\rm dyn}$ (c) and $\kappa\sigma^2_{\rm fc}$ (d) for nine (black points) and sixteen (red circles) centrality bins.}
\end{figure}

Using the corrected formulae Eq.~(5) and Eq.~(12) of ref. ~\cite{Luoxf-CBWC} and Eq.~(28) of ref. ~\cite{Luoxf-JPG39}, the corrected $\kappa\sigma^2$ for nine (black points) and sixteen (red circles) centrality bins are presented in Fig.~2(d). Where red and black points are close to each other. So $\kappa\sigma^2_{\rm fc}$ are centrality bin width independent. Formula corrected cumulants indeed eliminate the influence of centrality bin width.

At each of centrality bin, we construct corresponding mixed sample. Its $\kappa\sigma^2$ for nine (black solid points) and sixteen (red circles) centrality bins are presented in Fig.~2(b). Where influence of centrality bin width is similar to the case of the original sample, as showed in Fig.~2(a).  

The centrality dependence of dynamical cumulants ($\kappa\sigma^2_{\rm dyn}$) for nine (black points) and sixteen (red circles) centrality bins are presented in Fig.~2(c). The black solid points and red circles are close to each other. This shows that dynamical cumulants are centrality bin width independent, same as formula corrected cumulants in Fig.~2(d). This is understandable. Dynamical cumulants deduct the influence in the mixed sample, which is the same as the original sample.
 
It should be noted that dynamical cumulants in Fig.~2(c) is smaller than that of formula corrected cumulants in Fig.~2(d), but the same magnitude as those dynamical cumulants in Fig~1(d), where the statistical fluctuations are subtracted. Here, dynamical cumulants eliminate not only  statistical fluctuations, but also the influence of centrality bin width. While, formula corrected cumulants in Fig.~2(d) still contain statistical fluctuations. This is why dynamical cumulants are smaller than those of formula corrected.
  
For other cumulants of net-proton, same plots as those showed in Fig.~2 are not presented. Their dynamical cumulants are also centrality bin width independent, similar to $\kappa\sigma^2_{\rm dyn}$.

\subsection{Detection efficiency}

In heavy ion experiment, some particles are not detected. The detection efficiency described the percentage of detected particles. It usually depends on the transverse momentum, rapidity and azimuthal angule of particle, and is difficult to correct exactly. True distribution of particle number is somehow distorted by efficiency~\cite{Koch-PRC-efficiency,Koch-PRC91-027901}.

To study the influence of efficiency to cumulants, we simply take a fixed efficiency for all produced particles, i.e., randomly take $80\%$, or $60\%$, of produced particles. Centrality dependency of 
$\kappa\sigma^2$ of original sample for three values of efficiency, $100\%$ (black points), $80\%$(black squares), and  $60\%$ (blue triangles) are presented in Fig.~3(a), respectively. It shows that three kinds of points are separated. Cumulants vary with efficiency. The influence of efficiency is not negligiable. 

\begin{figure}
\includegraphics[width=1.0\linewidth,angle=-90]{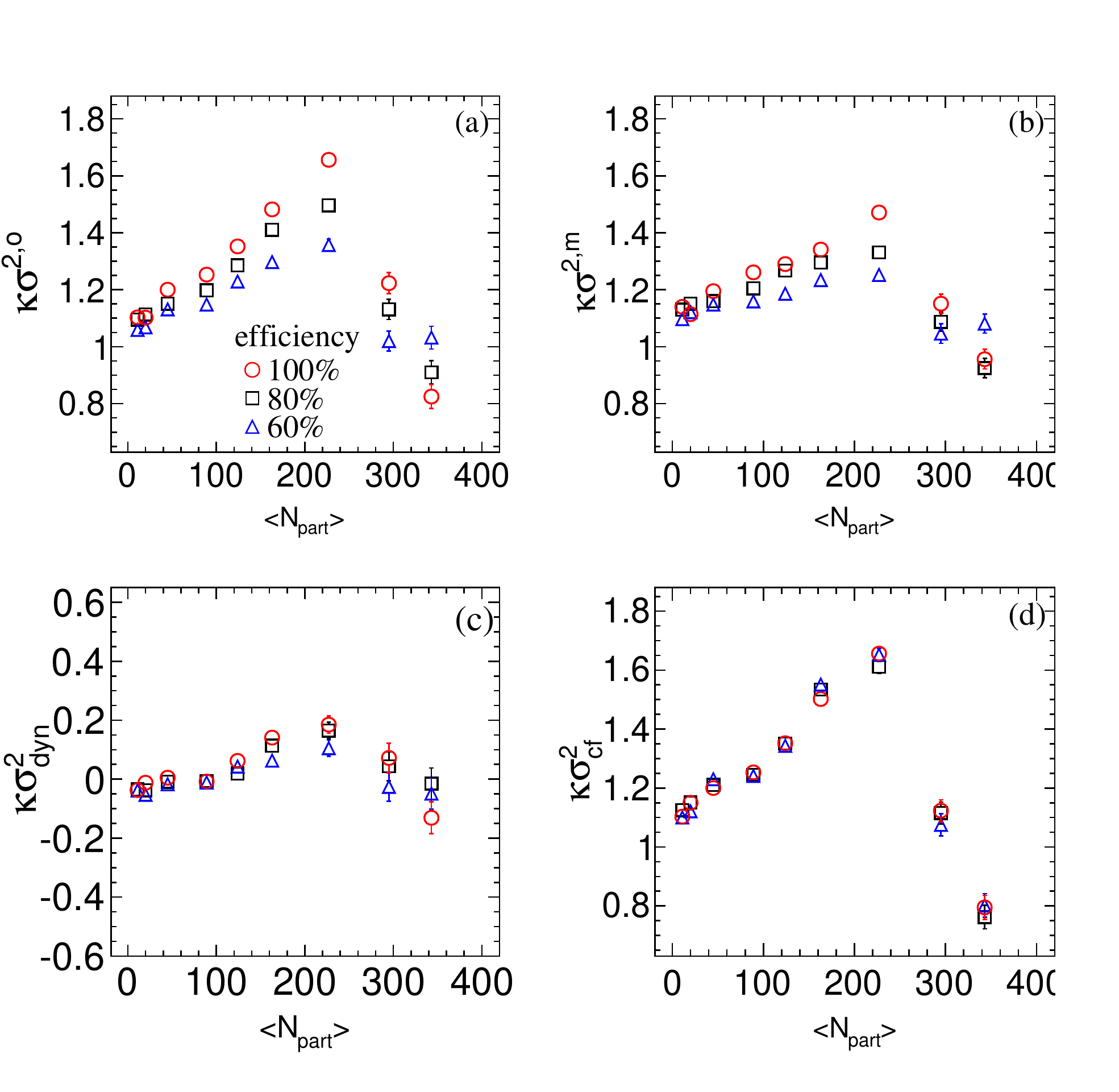}
\caption{\label{Fig. 3}The centrality dependency of $\kappa\sigma^2$ of original (a) and mixed (b) samples, dynamical (c) and formula corrected (d)$\kappa\sigma^2$ for three values of efficiency, $100\%$ (red circles), $80\%$(black squares), and  $60\%$ (blue triangles).}
\end{figure}

The centrality dependence of $\kappa\sigma^2$ of mixed samples for three values of efficiency, $100\%$ (red circles), $80\%$(black squares), and  $60\%$ (blue triangles) are presented in Fig.~3(b). Where the influence of efficiency are similar to that of the original sample as showed in Fig.~3(a). This is understable. The particles of mixed sample come from the original samples, and have exactly the same influence of efficiency.    

Using the formula Eq.~(26) of ref. ~\cite{Koch-PRC91-027901} and  Eqs.~(A20), (A23), (A25), and (A28) of ref. ~\cite{Luoxf-PRC91-034907}, centrality dependency of formula corrected $\kappa\sigma^2$ for three efficiencies, $100\%$ (red circles), $80\%$(black squares), and $60\%$ (blue triangles) are presented in Fig.~3(d). Where black squares and blue triangles overlap with those red circles within errors at each of centralities. So $\kappa\sigma^2_{\rm fc}$ is efficiency independent. 

The centrality dependency of dynamical $\kappa\sigma^2$ for three efficiencies, $100\%$ (black points), $80\%$(black squares), and $60\%$ (blue triangles) are presented in Fig.~3(c). Where three
kinds of points overlap within errors. So $\kappa\sigma^2_{\rm dyn}$ is also efficiency independent. This is because the mixed sample contains the same influence of efficiency as that of the original sample. Dynamical cumulants subtract the influence of efficiency remaining in the sample of mixed events.

Values of those points in Fig.~3(c) are the same magnitude as dynamical cumulants in Fig~1(d), where the statistical fluctuations are subtracted. So dynamical cumulants subtract not only the influence of detection efficiency, but also the statistical fluctuations, and the influence of centrality bin width.
While, formula corrected cumulants in Fig.~3(d) subtract only the influence of efficiency.

For other cumulants of net-proton, same plots as those showed in Fig.~2 are not presented. Their dynamical cumulants are also efficiency independent, similar to $\kappa\sigma^2_{\rm dyn}$.  

\section{Summary and conclusions}

In the paper, we first generate the sample of Au + Au collisions at 19.6 GeV by the AMPT default model. Then, at each of centrality bin, we construct its corresponding sample of mixed events, where multiplicity distribution keeps consistent with that of the original sample. Electic-charge particles of a mixed event are randomly taken from the pool, where all particles of the original events are put in. So global and systematic characters of the original sample remain in the mixed sample, and all concerning correlations inside an original event are lost. 

The centrality dependence of net-proton cumulants for the original and mixed samples is presented and compared. Dynamical cumulant is defined as the difference between cumulants of the original and mixed samples. It shows that dynamical cumulants are subtracted non-critical effects in the sample of mixed events. They are both centrality bin width and detection efficiency independent, in consistent with formula corrected cumulants. 

In conclusion, the mixed sample provides a good estimation for global and systematic effects. The defined dynamical cumulants substantially subtract influences of non-critical effects in the mixed sample. It helps us to study critical related fluctuations from higher cumulants of conserved charges at the RHIC BES I and II. 
 
\section{Acknowledgement}

This work is supported in part by the Ministry of Science and Technology (MoST) under grant No. 2016YFE0104800, and the Fundamental Research Funds for the Central Universities under
grant No. CCNU19ZN019.

\bibliographystyle{unsrt}
\bibliography{Subtracting}
\end{document}